\documentclass[11pt,oneside,letterpaper]{article}
\usepackage{amssymb}
\usepackage{amsmath}
\usepackage[dvips]{graphicx}
\usepackage{setspace}
\usepackage{amsfonts}
\usepackage{fancyhdr}
\usepackage{xcolor}
\usepackage{graphicx}
\usepackage{rotating}
\usepackage{comment}
\usepackage{color}
\usepackage{cite}
\usepackage{braket}
\usepackage{bbm}
\usepackage{float}
\usepackage{mathtools}
\usepackage{titlesec}
\definecolor{dolphingreen}{rgb}{0,0.74,0.63}
\definecolor{dolphinorange}{rgb}{0.98,0.53,0.098}
\definecolor{purple}{rgb}{0.4,.2,0.7}

\newcommand{\f}{\frac}
\newcommand{\Pf}{{\rm Pf}}

\usepackage[colorlinks=true,citecolor=dolphinorange,linkcolor=black,urlcolor=dolphingreen]{hyperref}
\usepackage{pdfsync}

\makeatletter
\newcommand*{\defeq}{\mathrel{\rlap{%
                     \raisebox{0.3ex}{$\m@th\cdot$}}%
                     \raisebox{-0.3ex}{$\m@th\cdot$}}%
                     =}
\makeatother

\DeclareMathOperator{\e}{\epsilon}

\def\be{\begin{eqnarray}}
\def\ee{\end{eqnarray}}

\newcommand{\<}{\langle}
\newcommand{\tr}{\textrm{Tr}\,}

\newcommand{\ttbar}{$T\bar{T}$ }
\newcommand{\bea}{\begin{eqnarray}}
\newcommand{\eea}{\end{eqnarray}}
\def\ben{\begin{equation}}
\def\een{\end{equation}}
\def\bne{\begin{equation}}
\def\ene{\end{equation}}

 \let\b=\beta \let\g=\gamma \let\d=\delta \let\e=\varepsilon
   \let\k=\kappa
\let\l=\lambda \let\m=\mu    \let\r=v
 \let\t=\tau

\let\w=\omega \let\G=\Gamma \let\D=\Delta

\let\f=\frac

\def\ba{\begin{array}}
\def\ea{\end{array}}
\def\del{\partial}

\def\vol{{\rm Vol}}

\titleformat{\paragraph}
{\normalfont\normalsize\bfseries}{\theparagraph}{1em}{}
\titlespacing*{\paragraph}
{0pt}{3.25ex plus 1ex minus .2ex}{1.5ex plus .2ex}

\allowdisplaybreaks  

\renewcommand*{\defeq}{\mathrel{\vcenter{\baselineskip0.5ex \lineskiplimit0pt
                     \hbox{\scriptsize.}\hbox{\scriptsize.}}}%
                     =}

\numberwithin{equation}{section}

\begin{document}
\onehalfspacing

\begin{center}
\begin{flushright}
BRX-TH-6659
\end{flushright}
~

\vskip5mm

{\LARGE  {\textsc{Hamiltonian deformations in quantum mechanics, \ttbar, and SYK}}}

\vskip10mm
David J. Gross$^1$, Jorrit Kruthoff$^2$, Andrew Rolph$^{1,3}$, and Edgar Shaghoulian$^4$

\vskip5mm

{\it 1)  Kavli Institute for Theoretical Physics,\\
University of California, Santa Barbara, CA 93106}\vskip3mm
{\it 2) Stanford Institute for Theoretical Physics, Stanford University, Stanford, CA 94305} \vskip3mm
{\it 3) Martin A. Fisher School of Physics,\\
Brandeis University, Waltham, MA 02453, USA}
\vskip3mm
{\it 4) Department of Physics, Cornell University, Ithaca, New York, USA }

\vskip5mm


\end{center}

\vspace{4mm}

\begin{abstract}
\noindent Motivated by $T \bar T$, we introduce and study a wide class of solvable deformations of quantum-mechanical theories. These deformations map the Hamiltonian to a function of itself. We solve these theories by computing all finite-temperature correlation functions of the deformed theory in terms of the correlators of the undeformed theory. Applications to AdS/CFT, SYK, and the Schwarzian theory are considered. We write down the deformed Schwarzian action for an arbitrary Hamiltonian deformation and find that the maximal Lyapunov exponent is unchanged. 
 \end{abstract}

\pagebreak
\pagestyle{plain}

\setcounter{tocdepth}{3}
{}
\vfill
\tableofcontents

\section{Introduction}\label{intro}
Calculable deformations of well-understood physical systems form the basis of much of theoretical physics. Often we have a simple theory with a handful of exact analytic solutions, and we try to learn about more diverse physical phenomena by deforming away from these special cases. A famous example is the three-body problem, where one tries to deform away from the Keplerian ellipses governing planetary motion to account for the influence of a third body. This approach was successful enough to predict the existence of Neptune in the 1800s.

Deforming away from well-understood systems is often also implemented in quantum field theory. We start from a theory which we can solve exactly, such as free particles, and introduce weak interactions. This framework is robust enough to explain phenomena ranging from the anomalous magnetic moment of the electron to  structure formation in the early universe.

Much of the recent progress in quantum field theory, however, has been spurred by exact nonperturbative techniques. In this paper we will introduce and study an infinite class of nonperturbative deformations to quantum field theories that can be solved exactly. These deformations map the Hamiltonian to a function of itself, $H\rightarrow f(H)$. In spacetime dimensions $d>1$, such deformations generically lead to nonlocal theories that break Lorentz invariance. This suggests considering the case $d=1$, i.e. ordinary quantum mechanics. As discussed in \cite{Anninos:2017hhn, Gross:2019ach}, this case is particularly well-suited to capture features of emergent spacetime due to the rich infrared. 

The integrability of these deformations relies crucially on the Hamiltonian being independent of time, so that it is a conserved charge. The class of deformations can be enlarged to mix in other conserved charges that appear in the theory being considered, as long as we pick a mutually commuting set. For example, we could deform the Hamiltonian of the hydrogen atom by a function $f(H,L^2, L_z)$. In this paper we will focus on functions of the Hamiltonian only, although many of the techniques discussed can be extended to the case with additional conserved charges. We will point out appropriate generalizations along the way.

An important feature of these deformations is that the energy spectrum of the deformed theory is known in terms of the energy spectrum of the undeformed theory, $E_i \rightarrow f(E_i)$. The other crucial feature is that the eigenvectors of the new and deformed Hamiltonian coincide (we will return to this point shortly). These two facts can be used in conjunction to write down formulas for the correlation functions of the deformed theory in terms of the correlation functions of the undeformed theory, which we will come to in the next section. 

We now clarify the claim that the eigenvectors of the deformed and undeformed Hamiltonians coincide. What is obviously true is that eigenvectors of $H$ remain eigenvectors of $f(H)$. For a finite-dimensional system, this is  sufficient to guarantee diagonalization of $f(H)$ by the eigenvectors of $H$, thus giving us the complete set of eigenvectors. This is true for arbitrary $f(H)$. For theories with an infinite-dimensional Hilbert space, like the simple harmonic oscillator, $f(H)$ can have new eigenvectors in general. To deal with this, we restrict to functions $f(H)$ that have an expansion parameter $\l$ with $f(H) \rightarrow H$ as $\l \rightarrow 0$. This means $f(H)$ can be interpreted as a deformation of our original system. Then we require, even if we work nonperturbatively in $\l$, that physical observables (and hence our eigenvectors) are analytic around $\l = 0$. This will enforce that the original eigenvectors are chosen. Dual to this picture is that of the Schr\"odinger equation for our deformed theory. Assuming we begin with a theory with canonical kinetic terms, our deformations do not induce higher derivative terms, although they do introduce higher powers of the kinetic term. This will in turn induce higher spatial derivatives in the Schr\"odinger equation, and our rule will imply additional boundary conditions that are chosen to restrict to the wavefunctions of the theory defined by $H$. This gives a perfectly well-defined model with  computable observables. As a close analogy, one can consider Dirac's classical theory of an electron in a background electromagnetic field \cite{Dirac:1938nz}. He wrote down an equation for the worldline of the particle that involved the third time derivative of its position, and to exclude unphysical states he imposed a future boundary condition (analogous to our spatial boundary condition to obtain the physical wavefunctions). Later, Bhabha pointed out that the unphysical solutions have a singular expansion in the electric charge $e$ and eliminated them by demanding smoothness as $e\rightarrow 0$ \cite{Bhabha:1946zz}. This is precisely what we do. In some instances, for example \eqref{spectralcurve}, we will restrict to strictly monotonic $f(H)$ to ensure the existence of $f^{-1}(H)$. This lets us avoid eigenvalue crossing.

\subsection*{\it Summary}
In section \ref{correlators}, we will present formulas for the correlation functions of the deformed theory in terms of the correlation functions of the undeformed theory. These formulas are integrals of the undeformed correlator against products of a kernel $K(\b, \b')$ which is defined so that $e^{-\b f(E)} = \int d\b' K(\b,\b') e^{-\beta' E}$.  A closed-form expression for the kernel is only available for special deformations (including the $1d$ \ttbar deformation introduced in \cite{Gross:2019ach} following \cite{Cavaglia:2016oda, Smirnov:2016lqw}), but it is straightforward to compute numerically in the more general case. Some closed form kernels are presented in appendix \ref{app:Kernels}, and some numerical calculations in a case without closed form kernels are presented in appendix \ref{numerics}. 

In section \ref{bulk} we will interpret these deformations in the context of AdS/CFT and show that they correspond to a new boundary value problem in the bulk which keeps fixed some combination of the metric and extrinsic curvature. For the $1d$ \ttbar deformation we show that the mixed boundary conditions at infinity can be re-interpreted as Dirichlet boundary conditions at some finite radius upon using the bulk equations of motion, as was shown in one higher dimension in \cite{Guica:2019nzm}.

Section \ref{SYK} is devoted to the Sachdev-Ye-Kitaev (SYK) model. We start by deforming the SYK model and consider both the $1d$ \ttbar deformation and the $f(H) = H + \l H^2$ deformation with $\l \sim 1/N$. After introducing the usual collective fields $G$ and $\Sigma$ we find that the solutions of the Schwinger-Dyson equations are changed by a renormalization $J \rightarrow J(\l)$. In the case where we shift the SYK ground state energy to zero before deforming the theory, the renormalization is trivial and the solutions remain unchanged. In a non-'t Hooft limit with $\l$ finite as $N\rightarrow \infty$, the effective action approach becomes difficult, but we can still obtain expressions for the deformed quantities using the integral transforms of section \ref{correlators}. 

We also consider first disorder averaging SYK and then deforming the resulting theory. This sequence of performing the disorder average first lets us solve for an arbitrary $f(H)$ deformation. We find that the solution to the deformed Schwinger-Dyson equations is again given as a renormalization $J \rightarrow J(\l_i)$ of the undeformed solutions, where $\l_i$ are the deformation parameters in $f(H)$. Again, when the ground state energy is shifted to zero, the deformation has no effect.

In section \ref{Schwarzian} we consider the Schwarzian and related theories and compute their deformed partition functions under a  class of $f(H)$ deformations in closed form. We also discuss arbitrary $f(H)$ deformations of the Schwarzian theory and are able to write down the deformed Schwarzian action explicitly. Upon computing fluctuations around the saddle of these theories and the OTOC, we find that the Lyapunov exponent remains maximal. 

\section{Correlation functions}\label{correlators}
In this section we show how to calculate correlation functions in the deformed theory from correlation functions in the undeformed theory.  We will start by reviewing and expanding the analysis of \cite{Gross:2019ach}, which just involves the thermal partition function. After that we move on to various correlators, building up slowly to $n$-point correlators.

\subsection{Thermal partition function}
The deformed partition function can be written as  
\be
Z(\b)_{\l} = \int_{-\infty}^{\infty} dE \,e^{-\b f(E)} \rho(E)= \int_{-\infty}^{\infty} dE\, e^{-\b E}\rho_{\l}(E) 
\ee
for which we immediately have
\be\label{spectralcurve}
\rho_{\l}(E) = \rho(f^{-1}(E)) \f{df^{-1}(E)}{dE}\,.
\ee
Here we are assuming a strictly monotonic $f(H)$ so that it is invertible. The deformed partition function can be written as an integral transform of the undeformed partition function by introducing a kernel defined as 
\be \label{eq:kernel}
K_f(\b,\b') = \f{1}{2\pi i}\int_{\mathcal{C}_0} dE e^{-\b f(E) + \b' E}
\ee
for some contour $\mathcal{C}_0$. This gives
\be
e^{-\b f(E)} = \int_{\mathcal{C}} d\b' e^{-\b' E} K_f(\b,\b') \implies Z(\b)_{\l} = \int_{\mathcal{C}} d\b' Z(\b') K_f(\b,\b')\,.
\ee 
We can give a simple expression for this kernel in a few cases: 
\be\label{holodeform}
f(H) = \f{1}{4\l}\left(1 - \sqrt{1-8\l H}\right) \implies  K_f(\b,\b') = \f{\b}{\b'^{3/2}\sqrt{-8\pi \l}}\exp \left(\f{(\b-\b')^2}{8\b' \l}\right)
\ee
\be\label{H2}
f(H) = H - 2 \l H^2 \implies K_f(\b,\b') = \f{1}{\sqrt{8\pi \l \b}} \exp \left(-\f{(\b'-\b)^2}{8\l\b}\right)
\ee
for $\l < 0$. These deformations are inverses of one other. The contour $\mathcal{C}$ of the first deformation runs from $0$ to $\infty$ and so the second one has a contour along the full imaginary axis with a small positive real part for convergence. The first deformation was proposed in \cite{Gross:2019ach} as the one-dimensional version of the $T\bar{T}$ deformation. We will consider various other deformations with explicit kernels in appendix \ref{app:Kernels}. Even without closed-form kernels, we can use \eqref{eq:kernel} to compute the kernel numerically. A simple example is provided in appendix \ref{numerics}. 


Besides integral transformations we can also introduce a differential operator whose action on the undeformed partition function gives the deformed partition function:
\be\label{diffop}
Z(\b)_\l =  \left(\sum_{i=0}^\infty\f{\b^{i}(-f(-\partial_\b)-\partial_{\b})^{i}}{i!}\right)Z(\b) =\vcentcolon \mathcal{D}_f (\beta) Z(\b)\,.
\ee
Each $-\partial_\b$ acts by bringing down a factor of $E_j$ from each Boltzmann factor, which is then manipulated into $-f(E_j)+E_j$ and re-exponentiated, i.e. $\mathcal{D}_f(\b)e^{-\b E_j} = e^{-\b f(E_j)}$. Notice, however, that the series in \eqref{diffop} is generically asymptotic and serves as a formal expression for the deformation. We will therefore focus on the integral transforms as our method of implementing the deformation. 

\subsection*{\it Additional conserved charges and the grand canonical ensemble}

The case where the deformation depends on additional conserved charges can be treated similarly, as long as all the additional conserved charges are mutually commuting. Let's consider the grand canonical ensemble with some chemical potentials $\m_i$ turned on.  The integral kernel will be a function of these additional potentials, $K_f(\b, \b', \m_i, \m_i')$, and the integral will be over $\b'$, $\m_i'$. Its role is still to transform Boltzmann factors $e^{-\b (E - \sum_i\m_i Q_i)}$ $\rightarrow e^{-\b (f(E, Q_i) - \sum_i \m_i Q_i)}$. 

\subsection*{\it One-point functions}
At the next level of complexity is a one-point function. Recall that we are not in a conformal theory so this need not vanish even at zero temperature. The operator can be placed at the origin by time-translation invariance.  The finite-temperature expectation value is given as 
\be
\sum_i e^{-\b f(E_i)}\langle E_i|O|E_i\rangle\,.
\ee
Notice that the expectation value $\langle E_i|O|E_i\rangle$ is the same between the deformed and undeformed theories since the eigenvectors $|E_i\rangle$ are unchanged under a deformation $H \rightarrow f(H)$. This is true as long as we keep fixed the operator $O$.\footnote{\label{footnoteO} We are stressing this since if one took $O$ to be the Hamiltonian $H$ then we would have to be careful when computing correlators, since what we call the Hamiltonian changes when we deform the theory.} This means that the same integral and differential transforms for the partition function apply to this case. The zero-temperature expectation value is simply $\langle 0|O|0\rangle$ and is unchanged between the two theories. Note that our one-point functions are normalized such that for $O$ being the identity operator, the one-point function is equal to the thermal partition function. In what follows we will continue to consider unnormalized correlation functions, which can be normalized by dividing by the deformed thermal partition function.

\subsection{General correlation functions}
 We will start with the vacuum two-point function as a simple, illustrative example. In the undeformed theory this can be written as 
\be\label{twopointEbasis}
\braket{O(\t)O(0)} = \sum_{i} e^{-E_i \t} |\braket{0|O|E_i}|^2,
\ee
where we will assume the ground state energy is zero and take $\t > 0$. The deformed correlator can be obtained by simply replacing the exponential by $e^{-f(E_i)\t}$, since the energy eigenvectors do not change under the deformation. As we saw in the previous subsection, such a change in exponential can be accomplished in two ways, either by an integral transform or by a differential operator. In fact, since we are again only changing a single exponential, the transformation from the undeformed to deformed quantity is the same as for the partition function and one-point function. 

Now we consider $n$-point thermal correlators.  Time ordering for simplicity, with $\t_1 > \cdots > \t_{n-1}>0$, we have
\begin{align}
C_0(\b,\{\t_i\}) &= \int dE_1 \langle E_1 |O(\t_1)\cdots O(\t_{n-1})O(0)|E_1\rangle e^{-\b E_1}\nonumber\\\label{formThermCor}
 &= \int \prod_{i=1}^{n} dE_i \langle E_1|O|E_2\rangle \cdots \langle E_{n}|O|E_1\rangle \exp\left(-\sum_{i=0}^{n-1}\b_iE_{i+1}\right)\,
\end{align}
where $\b_i = \t_i -\t_{i+1}$ for $i = 0,\dots, n-2$, $\t_0 = \b$ and $\b_{n-1} = \t_{n-1}$.

As long as there is a kernel to transform the partition function $Z(\b)_\l = \int d\b' K_f(\b,\b')Z(\b')$, we can tranform the exponentials in \eqref{formThermCor} using that kernel: 
\be
C_\l(\b,\{\t_i\}) = \int \left(\prod_{i=0}^{n-1} d\b'_i K_f(\b_i, \b_i')\right) C_0(\b,\{\t_i\})\,.
\ee
Again, these correlators can be canonically normalized by dividing by $Z_{\l}(\b)$. The formulas we derived here can also be applied to situations where there are additional conserved charges, with the only difference being the form of the differential operator and kernel. It is interesting to apply these formulas to seed Hamiltonians that are themselves integrable. Theories from undergraduate quantum mechanics, such as the harmonic oscillator or hydrogen atom, form a particularly fun set of examples, but there are also infinite classes of fancier Hamiltonians one could play with. These include, for example, supersymmetric partner Hamiltonians of exactly solvable systems \cite{Cooper:1994eh}, or Hamiltonians of the form $H = 2 \cosh p + V(x)$ for \emph{arbitrary} potential $V(x)$ \cite{Grassi:2018bci}. 

As a simple application let's consider correlators in the undeformed theory that have a definite scaling behaviour, for example in conformal quantum mechanics \cite{deAlfaro:1976vlx, Chamon:2011xk},
\be
\braket{O(\t)O(0)} = \frac{1}{\t^{2 \D}}\,.
\ee
The deformed correlator, in the case of the 1d \ttbar deformation \eqref{holodeform}, is given by
\be
\braket{O(\t)O(0)}_{\l} = \t^{-2\D +1/2} \frac{e^{-\t/(4\l)} K_{2\D + 1/2}\left(-\frac{\t}{4\l}\right)}{\sqrt{-2\pi \l}}\,.
\ee
For $\D = -1/2$ and $\D = 0$ the initial propagators are those of the free scalar and free fermion, respectively. In those cases  (and only those cases) the deformed correlator is identical to the undeformed one. For the free fermion this is simply because $H = 0$ and the deformation therefore does not do anything. For the free scalar it is a property of the worldline action resulting from the deformation (see section 3.2 of \cite{Gross:2019ach}), which is related to the representation of a free scalar in QFT$_d$ in terms of a worldline scalar.

For $\D>0$ the deformed correlator in the ultraviolet behaves as 
\be\label{deformedOshortdistance}
\braket{O(\t)O(0)}_{\l} = \frac{(-8\l)^{2\D}}{\t^{4\D}} \frac{\G(2\D + 1/2)}{\sqrt{\pi}} \quad \text{ as } \t \to 0\,.
\ee
The coincident divergence is still present in these correlators, but its nature is different due to the irrelevant deformation. The doubling of the conformal dimension at short times is the same as $\t \rightarrow \t^2$. This can be understood since small $\t$ corresponds to large energies, which in the deformed theory go as $\sqrt{E}$, with $E$ the original energy. 
Analogously, the quadratic deformation \eqref{H2} will halve the dimension $\Delta$. The evolution of correlation functions under the 2d $T\bar T$ deformation was studied in \cite{Cardy:2019qao}.

\subsection{Dispersion relation}\label{dispersion}
In this subsection we explore a different method of relating the deformed and undeformed two-point functions in Lorentzian signature. 

We work with the time-ordered Lorentzian vacuum two-point function, 
\bne \label{eq:GL} G(t) = i\bra{0} \psi(t) \psi(0) \theta (t) \pm \psi(0) \psi(t) \theta (-t) \ket{0}. \ene
$\psi$ may be bosonic or fermionic. Without loss of generality we assume the ground state energy to be zero. In frequency space the retarded propagator, which is the first term in \eqref{eq:GL}, is a sum of simple poles positioned at the energy eigenvalues of the system,
\bne \mathcal F [G^R](\omega) = \sum_n |c_n|^2 \left [ \frac{1}{ f(E_n)-\omega - i \epsilon} \right ]. \ene
where $c_n := \braket{0 | \psi |n}$. The only effect of our deformation $H \to f(H)$ is to move the position of the poles. Therefore all we need to find the propagator in the deformed theory are the positions of the poles and their residues in the undeformed theory. This is captured nicely by the spectral density, which is the imaginary part of the retarded propagator,
\bne \label{Img}
\text{Im}\,\mathcal F[G_0^R] (\omega) = \pi \sum_n |c_n|^2  \delta (\omega - E_n)\,.
\ene
Using this we write a dispersion relation for the retarded propagator in terms of the spectral density of the undeformed theory
\bne
\mathcal F [G^R](\omega) = \frac{1}{\pi} \int_{0}^{\infty} d\bar \omega  \frac{ \text{Im}\,\mathcal F [G_0^R] (\bar \omega)}{f(\bar \omega ) - \omega - i \epsilon}. 
\ene
As a simple check, consider the retarded propagator of a free massless fermion:
\bne \mathcal F[G^R_0] (\omega) = \frac{1}{2}\,\frac{1}{-\omega - i\epsilon}. \ene
This is unaffected by arbitrary $f(H)$ deformations, as it should be since $H = 0$ for the undeformed theory.

\section{AdS/CFT}\label{bulk}
\subsection{Deformations as mixed boundary conditions}
In the context of AdS/CFT it is simpler to work with the deformations at the level of the action. We begin with a general set of deformations where we deform the Lagrangian by some function of the stress tensor and the metric, 
\bne S \rightarrow S + \int d\t \sqrt{\g}\, M(T_{\t\t}, \gamma^{\t\t}).\ene
 We assume that the deformation depends on a parameter $\l$ such that $M\rightarrow 0$ as $\l \rightarrow 0$. There may be other dimensionful couplings $\l_i$. In AdS/CFT we often think of multitrace deformations that involve just the operator, but in this case we are also mixing in the source. 
 
In the undeformed theory, the variation of the action is
\be
\delta S = \f{1}{2}\int d\t \sqrt{\gamma}\, T_{\t\t}\delta \gamma^{\t\t}. 
\ee
This piece comes from the Gibbons-Hawking-York boundary term. The bulk term vanishes on the equations of motion. Dirichlet boundary conditions on the metric make this boundary term vanish too, leading to a well-defined variational principle. When we add the multitrace deformation to the one-dimensional action, we have to include it as a boundary term in the bulk theory. This leads to a new term in the variation of the action,
\be
\delta S = \f{1}{2}\int d\t \,\sqrt{\gamma}\left[\left(T_{\tau\tau}-\f{M}{\gamma^{\tau\tau}}+2\f{\partial M}{\partial \gamma^{\tau\tau}}\right)\delta \gamma^{\tau\tau} + 2\f{\partial M}{\partial T_{\tau\tau}} \delta T_{\tau\tau}\right].
\ee
We want to rewrite this in terms of a new operator $g(\gamma^{\tau\tau},T_{\tau\tau})$ and its source  $f(\gamma^{\tau\tau},T_{\tau\tau})$,
\be
\delta S = \f{1}{2}\int d\t \,g^{-1/2} f \delta g\,.
\ee
From this expression we can read off the new bulk boundary condition necessary for a well-posed variational problem. The deformation changes the boundary condition  from  holding $\gamma^{\t\t}$ fixed to holding $g (\gamma^{\tau\tau},T_{\tau\tau})$ fixed.
 
To find $f$ and $g$ we solve
\be 
g^{-1/2}f \f{\partial g}{\partial \gamma^{\t\t}} &= \sqrt{\gamma} \left(T_{\t\t}-\f{M}{\gamma^{\t\t}}+2\f{\partial M}{\partial \gamma^{\t\t}}\right),\quad 
g^{-1/2}f\cfrac{\partial g}{\partial T_{\t\t}} &= 2\sqrt{\gamma}  \cfrac{\partial M}{\partial T_{\t\t}}.
\ee
These may have several solutions; we make the additional restriction that $f \to T_{\t\t}$ and $g \to \g^{\t\t}$ as $\l \rightarrow 0$. 

\subsection{$f(T)$ deformations}
Let's consider deformations that are just a function of the trace of the stress tensor $T\defeq T_{\t\t}\g^{\t\t}$ and one parameter $\lambda$. For solutions to the variation of the action, we take the ansatz 
\bne f(\gamma^{\tau\tau}, T_{\tau\tau}) = T_{\tau\tau} F(T), \qquad g(\gamma^{\tau\tau}, T_{\tau\tau}) = \gamma^{\t\t} G(T). \ene These satisfy the restriction on solutions we made if $F(T)$, $G(T)\rightarrow 1$ as $\l \rightarrow 0$. With this ansatz, the above partial differential equations reduce to an ODE and an algebraic equation:
\be\label{bndryODE}
\partial_T \log G(T) = \frac{2 \partial_T M_{\l}}{T-M_{\l}},\quad \sqrt{G(T)} F(T) = 1 - \frac{M_{\l}}{T}\,.
\ee
The solution is found by integration subject to the boundary condition specified above. 

Consider $M_{\l}(T) = \l T^2$ as an example. The first equation in \eqref{bndryODE} has a divergence at $\l T = 1$, so we cannot integrate it beyond that. Notice that $T$ is like the Hamiltonian and thus a positive operator, so that this divergence only occurs for $\l > 0$. The solution is
\be\label{T2deform}
 f(\gamma^{\tau\tau}, T_{\tau\tau}) = T_{\t\t}(1 - \l T)^3, \quad g(\gamma^{\tau\tau}, T_{\tau\tau}) = \g^{\t\t}(1 - \l T)^{-4}.
\ee
The variational problem is modified from fixing the boundary metric to fixing a combination of the boundary metric and boundary stress tensor, $g(\g^{\t\t}, T_{\t\t})$.

\subsection{JT gravity at finite cutoff}
Now we discuss the deformation derived in \cite{Gross:2019ach}, which was proposed to correspond to a finite Dirichlet cutoff in AdS$_2$. To show that mixed boundary conditions at infinity correspond to Dirichlet conditions at finite cutoff we will eventually need to use the bulk equations of motion. Since we will go on-shell, for convenience we will consider a deformation that is equivalent on-shell,
\be \label{eq:OT_pert}
M_{\l} = -2\l O T\,.
\ee
Because of the appearance of the operator $O$, we need to generalize the above discussion to include the bulk dilaton.

We will add the term \eqref{eq:OT_pert} perturbatively to the existing action, which can be the boundary action found after flowing with respect to the same deformation for some amount. Our task is now to find deformed metric, dilaton, stress tensor and $O(\l)$ such that
\begin{align}\label{holoDef}
\int d\t \sqrt{\g(\l)}\left(\frac{1}{2}T_{\t\t}(\l)\d\g^{\t\t}(\l) + O(\l)\d\Phi_b(\l)\right) = \int d\t \sqrt{\g}&\left(\frac{1}{2} T_{\t\t}\d\g^{\t\t} + O\d\Phi_b \right) \nonumber\\
&\quad\quad -2 \d \left(\int d\t \l\sqrt{\g}OT\right),
\end{align}
where $\Phi_b$ refers to the boundary value of the bulk dilaton. It is easily checked that 
\be \begin{split} \label{eq:hf}
\g_{\t\t}(\l) &= \g_{\t\t}(1 + 2 \l O)^2, \qquad \Phi_b(\l) = \Phi_b -2 \l T,  \\
T_{\t\t}(\l) &= T_{\t\t}(1 + 2 \l O)^2, \qquad  O(\l) = \frac{O}{1 + 2 \l O},\\
\end{split} \ee
form a solution for the deformed quantities. 

Note that the starting operator is the same as the deformed one, 
\bne  O T \sqrt{\g} =O(\l)T(\l)\sqrt{\g(\l)}, \ene
thus the above solutions are correct to all orders in $\l$. This solution can also be found by analyzing a first order flow as in \cite{Guica:2019nzm}.

\subsection*{\it Bulk analysis}
To show the equivalence of \eqref{eq:hf} to Dirichlet boundary conditions at a finite radial position, we need to relate the quantities above to bulk gravitational variables and use the equations of motion.  For JT gravity, these variables are the metric $g_{\mu\nu}$, dilaton $\Phi$, the trace of the extrinsic curvature $K$ (of the radial slice) and the normal derivative (to the radial slice) $\partial_n \Phi$. Let us work in Fefferman-Graham gauge,
\be
ds^2 = \frac{dr^2}{r^2} + r^2 \gamma_{\tau\tau} (r,\tau) d\t^2.
\ee
The conformal boundary is at $r \to \infty$. AdS$_2$ boundary conditions require $\lim_{r \to \infty} \g_{\t\t}(r,\t) = \g_{\t\t}^{(0)}(\t)$. In vacuum AdS$_3$ we know that the Fefferman-Graham expansion truncates at order $1/r^4$. Since JT gravity is a dimensional reduction of Einstein gravity in AdS$_3$, the expansion also truncates at order $1/r^4$ in this case.  The constraint $R = -2$ determines $\g^{(4)}_{\t\t}$ in terms of $\g^{(0)}_{\t\t}$ and $\g^{(2)}_{\t\t}$, and we have 
\be \label{solgFG}
\g_{\t\t}(r,\t) = \g_{\t\t}^{(0)}(\t) \left( 1+ \frac{1}{2r^2}\frac{\g^{(2)}_{\t\t}(\t)}{\g_{\t\t}^{(0)}(\t)} \right )^2. 
\ee
The dilaton can be written as 
\be\label{solphiFG}
\Phi(r,\t) = r \left( \Phi^{(0)}(\t) + \frac{\Phi^{(1)}(\t)}{r^2}\right),
\ee
with $\Phi^{(i)}(\t)$ determined in terms of $\g_{\tau\t}$ through the metric equation of motion. 

To convert to field theory variables we apply the same method as in \cite{Gross:2019ach, Hartman:2018tkw}: $\g_{\t\t}^{(0)}$ is the field theory metric and $\Phi^{(0)}$ is the source for the operator dual to the dilaton. From the renormalized on-shell action
\bne S_{E} = - \f{1}{2\kappa^2}\int_M d^2 x \sqrt{g}\Phi (R-2) - \f{1}{\kappa^2} \int_{\del M}  \Phi (K-1) \ene
the stress-tensor and operator $O$ are obtained by taking functional derivatives with respect to their sources, $\Phi^{(0)}= \lim_{r\to \infty} \Phi/r$ and $\g^{(0)}_{\t\t} = \lim_{r\to \infty} g_{\t\t}/r^2$. Explicitly, 
\begin{align}
O &=  \frac{1}{\sqrt{\g^{(0)}}}\frac{\d S_{E}}{\d \Phi^{(0)}}  = -\lim_{r\to \infty} \frac{r^2}{\k^2} (K - 1)\,,\\
 T_{\t\t} &=  -\frac{2}{\sqrt{\g^{(0)}}} \frac{\delta S_E}{\delta \g^{(0)\t\t}}   = -\lim_{r \to \infty}\frac{g_{\t\t}}{\k^2 r}\left(1 - n^r\partial_r\right) \Phi  \,.
\end{align}
To take the limit for $O$ we first rewrite $K$ by taking the divergence of the normal vector $n^{\mu} =r \d^{r\mu}$. This leads to
\be\label{Ograv}
O = \f{1}{\k^2}\f{\g_{\t\t}^{(2)}}{\g_{\t\t}^{(0)}}\,,
\ee
while for the stress tensor $T_{\t\t}$, using \eqref{solphiFG}, we find
\be\label{Tgrav}
T_{\t\t} = -\frac{2 \Phi^{(1)}(\t) \g_{\t\t}^{(0)}}{\k^2}\,.
\ee
\subsection*{\it Dirichlet boundary condition at finite radius}
We can now find what boundary condition the $-2\lambda OT$ deformation imposes on the gravitational quantities. The quantities to be held fixed are given in the first line of \eqref{eq:hf}. Upon translating to their gravitational duals using \eqref{Ograv} and \eqref{Tgrav}, we find that we want to keep the following fixed, 
\be\label{holosol}
\g_{\t\t}(\l) = \g_{\t\t}^{(0)}\left( 1 + \frac{2\l}{\k^2} \frac{\g^{(2)}_{\t\t}(\t)}{\g_{\t\t}^{(0)}(\t)}  \right)^2, \quad \Phi(\l) = \Phi^{(0)} + \frac{4\l}{\k^2}\Phi^{(1)}.
\ee
One can also write down expressions for $T_{\t\t}$ and $O$ in the deformed theory. Comparing \eqref{holosol} to the Fefferman-Graham expansions \eqref{solgFG} and \eqref{solphiFG} (upon performing the usual rescaling of the dilaton to go back to bulk variables \cite{Gross:2019ach}), we see that this particular boundary condition corresponds to a Dirichlet boundary condition at finite cutoff $r = r_c$, provided we identify
\be
\frac{4\l}{\k^2} = \frac{1}{r_c^2}
\ee
with $\l > 0$. This is precisely the dictionary advocated in \cite{Gross:2019ach}.

What about the deformation written purely in terms of $T$ as in \cite{Gross:2019ach}? Naively, this only sets up Dirichlet boundary conditions for the metric, since the dilaton is not involved in the deformation. But as we saw above, to show the equivalence with a finite Dirichlet cutoff we had to go on-shell. Going on-shell means the metric and dilaton are mixed, so this form of the deformation will also correspond to Dirichlet boundary conditions for the dilaton at some finite radius. Much of the analysis of \cite{Gross:2019ach} was done in a different gauge, with $\Phi^{(1)} = 0$ but nontrivial radial lapse $g_{rr}$. So the analysis above should be done in that gauge, although the end result will be the same. 

\section{Sachdev-Ye-Kitaev model}\label{SYK}
It is natural to consider the application of the techniques developed above to particular quantum-mechanical theories, like the harmonic oscillator or the hydrogen atom. Here we will consider a different theory, the SYK model.

The undeformed SYK Hamiltonian is \cite{kitaevKITP, PhysRevLett.70.3339}
\be
H = i^{q/2} \sum_{{i_j}} J_{i_1\cdots i_q} \psi_{i_1}\cdots \psi_{i_q},
\ee
where $J_{i_1\cdots i_q}$ is drawn from a random gaussian distribution with zero mean, but variance
\be\label{Jtwopoint}
\braket{J_{i_1\cdots i_q}^2} = \frac{J^2(q-1)!}{N^{q-1}}.
\ee
The notation $J_{i_1\cdots i_q}$ is rather cumbersome and so in what follows we will abbreviate it with $J_A$ where $A$ is a multi-index.

We will consider the SYK model with a shift in the Hamiltonian, $H - E_0$ for some constant $E_0$. The SYK model has a negative ground state energy, so we can tune $E_0$ to normalize the ground state energy to zero, which we have been assuming in previous sections. A constant shift has a trivial effect on the undeformed theory, however, once we deform the Hamiltonian to a function of itself, this choice becomes important and gives inequivalent deformed theories. 

We consider two ways of deforming the SYK model. We can either disorder average then deform the Hamiltonian, or vice versa. We begin by deforming the microscopic SYK Hamiltonian and then disorder averaging it. For general functions of the Hamiltonian it is not possible to integrate over couplings in the disorder average. It can be done for simple deformations like $f(H) = H + \l H^2$ and the 1d \ttbar deformation. The second situation amounts to first doing the disorder average and then deforming it. In this case we can define a Hamiltonian of the disorder averaged theory and consider general deformations $f(H)$. While deforming and then disorder averaging provides a microscopic picture of the physics, we can treat many more deformations if we disorder average and then deform, and we will see that the physics of the two cases where we understand them is similar. 

The integral transforms of section \ref{correlators} can be applied to the SYK model as long as we normalize the vacuum energy to zero. Picking a 't Hooft like scaling where $\l \sim N^{-1}$, we can immediately see that the vacuum two-point function, which scales as $N^0$, will be unchanged. This is because the kernel localizes to a delta function as $\l \rightarrow 0$. In this case, since the kernel is being multiplied by an object with no $N$-dependence, scaling $\l \sim N^{-1}$ and taking $N\rightarrow \infty$ is sufficient to localize the kernel to a delta function. We will corroborate this expectation below with explicit computations of the vacuum two-point function obtained from an effective action of collective fields. 

\subsection{Deforming before averaging: quadratic deformation}
In this section we derive the effective action of the deformed theory when we first deform and then disorder average. We will consider the deformation $f(H) = (H-E_0) + \l (H-E_0)^2$, where $E_0$ is some constant shift to the undeformed theory. First we compute the annealed disorder average, which is obtained by treating the random coupling $J_A$ as a scalar with a two-point function whose non-zero part is given by \eqref{Jtwopoint}. To get the disorder averaged partition function we integrate over $J_A$: 
\bne 
\< Z \rangle_J \sim \int dJ_{A} \exp \left(- \frac{N^{q-1}}{2 J^2 (q-1)!} J_{A}^2 \right) Z(J_A). 
\ene
The deformed Hamiltonian is
\bne
H_{\l} = \left(\sum_A J_A \psi^q_A (\tau)- E_0 \right) + \l \left(\sum_A J_A \psi^q_A(\tau) - E_0\right)^2.
\ene
Note that if multi-indices $A$ and $B$ have any index in common then $\psi_{AB}^{2q} = 0$, as the square of a Grassmann variable is zero. 

We are now ready to compute the $G, \Sigma$ effective action of this deformed theory. Just like the undeformed theory, this is done by disorder averaging over the couplings, inserting a resolution of identity involving the bilocal field $G(\tau,\tau')$, and finally integrating out the fermionic field $\psi$. 
We drop the constant factor $-E_0 + \l E_0^2$ in the action and omit the explicit time dependence of the fermions to simplify notation. The disorder averaged partition function then reads, 
\begin{align} \< Z \rangle_J  
=\int D\psi &\prod_A dJ_A \exp   \bigg[-\frac{1}{2}\sum_i \int \psi_i \partial_\t \psi_i - \nonumber\\
&\sum_{B,C}\bigg(\frac{1}{2\< J^2 \rangle} \delta_{B,C} + \l \int \psi^{2q}_{BC} \bigg)J_B J_C    
 - (1-2\l E_0) \sum_B J_B \int \psi_B^{q}\bigg].
 \end{align}
Just like the undeformed SYK model, at this stage we have a Gaussian integral in the couplings $J_A$. This is a consequence of the $H + \lambda H^2$ deformation being quadratic; the integral is not Gaussian for more general deformations. Unlike undeformed SYK, the determinant prefactor arising from the Gaussian integral is not a constant. The prefactor depends on $\psi$, so it needs to be exponentiated and included in the action:
\begin{align} \label{eq:partition function}
&\< Z \rangle_J = \int  D\psi \exp \bigg (-\frac{1}{2} \sum_i \int \psi_i \partial_\t \psi_i - \frac{1}{2} \tr \log (1 + Y) +\frac{1}{2} (1-2\l E_0)^2 \langle J^2 \rangle\tr [(1+Y)^{-1} X]\bigg ) 
\end{align}
where
\begin{align}
X_{AB} := \int \psi_A^q \int \psi_B^q \qquad
Y_{AB} := 2\lambda \langle J^2 \rangle \int \psi_{AB}^{2q}\,.
\end{align}
One way to proceed is to expand $\log (1+ Y)$ in powers of $Y$, and $(1+Y)^{-1} X$ in powers of $Y^n X$, then take the trace. This gives two expansions with an infinite number of terms. Each term  is a product of fermionic fields which can be replaced with powers of the bilocal field $G_\l(\tau, \tau')$. The result for the effective action is
\begin{figure}[t]
\centering
\includegraphics[width=15cm]{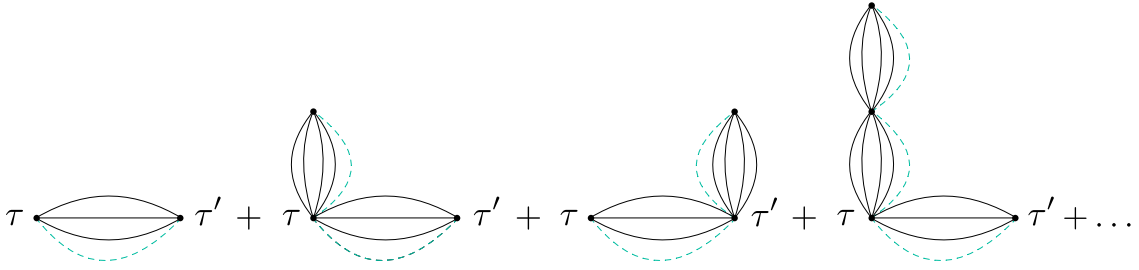}
\caption{\label{DiagramSigma}The leading order diagrams that contribute to $\Sigma (\tau , \tau')$ when $\lambda \sim 1/N$. Unlabelled vertices are integrated over. Chains formed of links are made of $q$ vertices. Each additional link in the chain adds a factor of $\lambda N$. The aqua blue dashed line indicates the disorder average.}
\end{figure}
\begin{align}\label{deformedI}
&\frac{I[G_\l,\Sigma_\l]}{N} = - \log \Pf (\del_\tau - \Sigma_\l) + \frac{1}{2} \int d\tau d\tau' \Sigma_\l (\tau, \tau') G_\l(\tau, \tau')
\nonumber\\ & +\frac{1}{2} \sum_{n=2}^\infty \frac{(-2\lambda N J^2/q)^n}{nN}\int G^q (\tau_1 , \tau_2)... G^q(\tau_n,\tau_1) 
\nonumber \\ &-\frac{1}{2}\frac{J^2 (1-2\l E_0)^2}{q} \sum_{n=0}^\infty \left(-\frac{2\l NJ^2}{q}\right)^n \int d\tau_1 ... d\tau_{n+2} G^q (\tau_1,\tau_2)... G^q (\tau_{n+1}, \tau_{n+2}).
\end{align}
We will now scale $\l \sim 1/N$. Keeping $\l$ order one will be discussed momentarily. The two infinite sums have a simple diagrammatic interpretation. Each term in each of the sums is represented by a chain diagram containing $n$ links, with each link containing $q$ edges. In the first sum the chain closes into a loop, in the other it does not. Since $\l\sim 1/N$, the first sum in \eqref{deformedI} is subleading, whereas the second remains finite.

The Schwinger-Dyson equations are obtained from the deformed action by taking functional derivatives with respect to $G$ and $\Sigma$. The  $\Sigma$ equation is unchanged from the undeformed theory and still arises from a geometric sum of 1PI diagrams,
\begin{align} \label{eq:SD2}
\delta(\tau) = \del_\tau G(\tau) - \int d\tau' G(\tau-\tau') \Sigma(\tau')\,.
\end{align}
Diagrammatically, taking a functional derivative of the action with respect to $G$ to calculate the self-energy removes a single propagator from each diagram in the two infinite families of open and closed chain diagrams. The resulting diagrams contributing at leading order are depicted in figure~\ref{DiagramSigma}.

Assuming translation invariance of $G(\tau, \tau')$, the second of the infinite sums can be written as powers of $\int d\tau G^q (\tau)$ and resummed. The end result for the self-energy has a perfectly well-behaved $N \to \infty$ limit,
\begin{align} \label{eqn:self_energy}
\Sigma(\tau, \tau')
&=  \left (\frac{1-2\l E_0}{1+\frac{2\l N J^2}{q}\int d\tau'' G^q (\tau'')}\right)^2  J^2 G^{q-1} (\tau, \tau') \,.
\end{align}
This equation is identical to the undeformed case, except for a renormalization of the coupling. The solution $G$ must have the same functional form as $G_0$, with a renormalized $J \to J(\lambda)$. (There is a self-consistency relation since $J(\l)$ depends on $G$ which depends on $J(\l)$. We will address this in section \ref{sec:4.3}.) One can check, by plugging the ansatz $G(J \tau) = G_0 (J(\l) \tau)$ into the Schwinger-Dyson equations and tuning $E_0$ to the vacuum energy, that the only consistent solution is $J(\l) = J$ at zero temperature. So the vacuum correlator is unchanged. At finite temperature the solution will change.

The invariance of the vacuum two-point function was explained at the beginning of this section from the point of view of the integral transforms of section \ref{correlators}. Let's look at an explicit example, the large-$q$ vacuum two-point function of SYK ($\t >0$),
\bne G_0 (\tau) = \frac{1}{2} \frac{1}{(1 + \mathcal{J} \tau )^{2/q}} , \qquad \mathcal J := \frac{\sqrt{q}J}{2^{\frac{q-1}{2}}}\,. \ene
Applying the integral transform \eqref{H2} gives the two-point function,
\bne \label{eq:M} G (\tau) = \frac{1}{2}\frac{1}{(2\mathcal{J}^2 \l \t)^{1/q}} U \left(\frac{1}{q}, \frac{1}{2}, \frac{(1+\mathcal{J}\t)^2}{4\l \mathcal{J}^2 \t} \right),
\ene
where $U$ is the confluent hypergeometric function of the second kind. For $\lambda \sim N^{-1}$ and smaller, $G(\t) = G_0(\t)$ at leading order. This agrees with what we found from the effective action: the deformation has no effect on the two-point function when $\lambda \sim N^{-1}$. 

Interestingly, the above two-point function remains finite when $\lambda\sim O(1)$. This is not special to large $q$. As neither the SYK correlation functions nor the integral transforms have any explicit dependence on $N$, neither do the deformed correlation functions. Keeping $\lambda$ finite as $N \rightarrow \infty$ is especially interesting and in other contexts corresponds to an M-theory limit (see e.g. \cite{Herzog:2010hf} for some field theory calculations in such a limit). These limits are much harder to study than standard 't Hooft type limits and often require supersymmetry to compute certain observables, but in our case the integral transforms from section \ref{correlators} can easily handle such a limit. It would be interesting to reproduce these correlators from an effective action point of view, the way we have done for $\l \sim N^{-1}$. 
 
\subsection{Deforming before averaging: 1d \ttbar deformation}
More general deformations will introduce higher powers of the disorder, rendering the disorder integral impossible to perform. However, for the $1d$ \ttbar deformation we can employ a trick. Call the starting Hamiltonian $H_0$. The deformed Hamiltonian and (Euclidean) Lagrangian are
\begin{align}\label{orig}
H(\l) &= \frac{1}{4\l}\left( 1 - \sqrt{1-8\l (H_0-E_0)} \right),\\
L_E(\l) &= \psi_i \partial_\t \psi_i + \frac{1}{4\l}\left( 1 - \sqrt{1-8\l (H_0-E_0)} \right).
\end{align}
At this stage it is hard to do the disorder averaging, since $H_0$ appears inside the root. But we can linearize the appearance of $H_0$ at the expense of introducing another field, which can be interpreted as an einbein. This is similar to the Polyakov trick for rewriting the Nambu-Goto action, and we have 
\be\label{ngtrick}
S_E(\l) = \int d\t e \left( e^{-1}\f 1 2\psi_i \partial_\t \psi_i - \frac{1}{8\l }(1 - e^{-1})^2 + H_0-E_0 \right).
\ee
Integrating out $e$ by picking the positive root (einbeins need be positive) gives us the action in \eqref{orig}. The  reason this was possible is that the potential in \eqref{orig} is a root of a quadratic equation:
\be
H(\l) + H_0-E_0 - 2 \l H(\l)^2 = 0\,.
\ee
We can therefore write an action
\be
S_E(\l) = \int d\t \left(\f 1 2 \psi_i \partial_\t \psi_i + X - e(X-H_0+E_0 - 2 \l X^2)\right),
\ee
where we introduced a field $X$ and a Lagrange multiplier $e$ enforcing the quadratic constraint which will identify $X = H(\l)$. For $\l < 0$, restricting to positive $X$ picks out the appropriate root of the quadratic constraint. Integrating out $X$ puts the action in the form \eqref{ngtrick} where we can interpret $e$ as an einbein. Notice that this action is exactly our worldline gravity action found in \cite{Gross:2019ach} in static gauge. 

Using this action, we can perform the disorder average easily. Let us consider the SYK theory as our initial Hamiltonian. The disorder averaged action is then
\be
S_E(\l,e) = \int d\t \left( \psi_i \partial_\t \psi_i - \frac{e}{8\l }(1 - e^{-1})^2-eE_0\right) - \frac{N}{2q}\int d\t d\t' J^2 e(\t)e(\t') G(\t,\t')^q .
\ee
The path integral over einbeins can be thought of as making $J$ dynamical, i.e. $J^2 e(\t) e(\t')\rightarrow J(\t,\t')^2$. Notice that again we need to scale $\l \to \l/N$ in order to have a 't Hooft large $N$ limit. From here we can introduce the bilocal field $\Sigma$ and integrate out the fermions. The Schwinger-Dyson equations for $G$ and $\Sigma$ take the usual form, but now with $J^2$ replaced by $J^2e(\t)e(\t')$. The $e$-equation of motion takes the form
\be\label{eeom}
\frac{e^{-2}-1}{8\l} - \frac{J^2}{q} \int d\t e(\t) G(\t,\t')^q-\frac{E_0}{N} = 0.
\ee
Since $f(H)$ deformations do not break any symmetries of the original theory, we will insist on translationally symmetric solutions. This means $e(\t)$ is independent of $\t$ and can be pulled out of the above integrals. As a result, the solutions for $G$ and $\Sigma$ will remain the same, but now have a renormalized $J \to J e$, similar to what we found previously. The only thing we need to know now is the integral of $G^q$, which by dimensional analysis takes the form $c/(Je)$ for some constant $c$. We therefore find
\be
e^{-1} = \sqrt{1 + 8\l \left(\frac{E_0}{N}+\frac{ c J }{q}\right)}
\ee
as a solution to \eqref{eeom}. Notice that when $E_0$ is set to equal the vacuum energy of the undeformed theory $-c N J/q$ we find $e = 1$, so there is no change in the deformed correlator $G$. This is consistent with the general argument in the beginning of this section using the integral transform of section \ref{correlators}. Note that we can again apply the integral transforms (in this case \eqref{holodeform}) and obtain results for the deformed two-point function for $\lambda \sim O(1)$.  
\subsection{Deforming after averaging}\label{sec:4.3}
In this subsection we will consider deformations of SYK \emph{after} performing the disorder average. This is a slightly unusual thing to do but the physics of the resulting system is similar to deforming first and then disorder averaging.  We start with the following undeformed action, where we shift by a constant to accommodate the case of subtracting the vacuum energy:
\be
S_E = \int d\t\left(\frac{1}{2} \psi_i \partial_\t\psi_i - i^q \frac{(q-1)!J^2}{2 N^{q-1}}\int  d\t'\,\psi_{A}(\t) \psi_A(\t')-\f{E_0}{2}\right)\,,
\ee
where $\psi_{A}(\t) = \psi_i(\t) \cdots \psi_{i_q}(\t)$ as before. We will return to this factor of $1/2$ in the $E_0$ shift momentarily. Written in this form, the Hamiltonian (generator of $\t$ translations) is
\be
H = - i^{q}\frac{(q-1)!J^2}{2 N^{q-1}}\int d\t' \psi_{A}(\t)\psi_{A}(\t')-\f{E_0}{2}.
\ee
To keep a conventional large-$N$ limit, we consider deformations of the form
\bne H \to N f(H/N). \ene
 It is straightforward to introduce the collective variables $G$ and $\Sigma$ in the usual way, giving a deformed action
\be
S_{E,\l} = N \left( - \log \Pf(\partial_\t - \Sigma) + \frac{1}{2} \int d\t \left[\int d\t' \Sigma(\t,\t')G(\t,\t') + 2f(H/N) \right] \right),
\ee
with
\be \label{eq:HG}
H =-  \frac{J^2 N}{2q}\int d\t' G(\t,\t')^q-\f{E_0}{2}.
\ee
Now we understand the factor of $1/2$, since if we tune $E_0$ to equal the vacuum energy of the undeformed theory, we see that this Hamiltonian is bounded below by zero due to the relation $\f{NJ^2}{q}\int d\t G_0^q = -E_0$ in the undeformed theory. The energy defined this way differs by a factor of two from computing $- \partial_\beta \log Z$; this difference is unimportant.

The Schwinger-Dyson equations are
\begin{align}\label{SD1}
\int d\t' G(\t,\t')\Sigma(\t',\t'') - \partial_{\t}G(\t,\t'') = -\d(\t-\t'')\,,\\
\Sigma(\t,\t') - f'(H/N ) J^2 G^{q-1}(\t,\t') = 0 \,.
\end{align}
Solving the second equation for $\Sigma$ and plugging into the first equation gives
\be
 J^2 f'(H/N)\int d\t' G(\t,\t')G(\t',\t'')^{q-1} - \partial_{\t}G(\t,\t'') = - \d(\t-\t'')\,.
\ee
This equation seems rather difficult to solve, because of all the $\int G^q$ factors that can appear in $f'(H/N)$, but it is formally the same as the undeformed equations if we identify 
\bne \label{eq:Jl} J(\l)^2 = J^2 f'(H/N). \ene 
Our proposed solution to the Schwinger-Dyson equations is 
\bne \label{eq:Gansatz}G(\t,\t') = G_0(\t,\t'; J(\l)), \ene where we take the undeformed correlator and map $J\rightarrow J(\l)$. $J(\lambda)$ is given by the solution to \eqref{eq:Jl} which is smoothly connected to $J$. Equation \eqref{eq:Jl} as written is a self-consistency relation, since $J(\l)$ depends on $H$ which depends on $G$ which depends on $J(\l)$, although we will see below how it can be recast as an algebraic relation for which we should expect solutions. 


This equation can be simplified. With our ansatz for the deformed two-point function \eqref{eq:Gansatz}, assuming zero temperature and using dimensional analysis fixes
\be\label{Gintdeform}
\int d\t \,G_{0}^q(\t; J(\l)) = \f{c}{J(\l)}.
\ee
$c$ is some dimensionless constant which depends on $q$.
Our equation to solve \eqref{eq:Jl} for the renormalized $J(\lambda)$ then becomes
\be\label{selfcon2}
J(\l)^2 =J^2 f' \left(-  \frac{J^2 c}{2q J(\l)}-\f{E_0}{2N} \right).
\ee
This is an algebraic relation for $J(\l)$. Note that if we tune $E_0$ to the vacuum energy of the undeformed theory, then the undeformed correlator -- without any renormalization of $J$ -- serves as a solution to the deformed equations of motion. This is because the undeformed correlator satisfies $\f{NJ^2}{q}\int d\t G_0^q = -E_0$. 

Now that we know the general picture, let us consider some examples where we can find the deformed correlators explicitly. 
\subsection*{\it An example: $q = 2$ SYK}
Let us consider the $1d$ \ttbar deformation $f(x) = (1- \sqrt{1- 8\lambda x})/(4\lambda)$. We know already that replacing $J$ with $J(\l)$ is a solution to the deformed Schwinger-Dyson equations, so the only thing left to do is find $J(\lambda)$ by solving \eqref{selfcon2}. To do so, we first need to calculate the constant of proportionality $c$ in \eqref{Gintdeform}. Our ansatz for the deformed propagator is given in terms of the undeformed $q=2$ propagator,
\bne  G(\tau) = G_0 (\tau; J(\lambda)) = \text{sgn}(\tau) \int_0^\pi \frac{d\theta}{\pi} \cos^2 \theta e^{-2 J(\lambda) |\tau | \sin \theta}. \ene
To find $c$ we calculate
\bne \int d\tau G^q (\tau) = -\frac{\del_\tau G(\tau)|_{\tau \to 0^+}}{J(\lambda)^2} = \frac{4}{3\pi J(\lambda) } \ene
giving $c = 4/3\pi$. The first equality above follows from the Schwinger-Dyson equations. The equation to solve for $J(\l)$ is
\be
\sqrt{1 +\frac{8 \l J^2}{3\pi J(\l)}} = \frac{J^2}{J(\l)^2},
\ee
where we set $E_0=0$. There are four solutions, and $J(\l)$ is fixed by demanding $J(0) = J$.

We can study the density of states to see what happens to the IR and UV for both $\l > 0$ and $\l < 0$. The density of states is simply the Wigner semi-circle \cite{10.2307/1970079}, but now with a $\l$-dependent $J$:
\be
\rho(E) = \frac{1}{J(\l)}\sqrt{1-\left(\frac{E}{2J(\l)}\right)^2}.
\ee
The IR value is $1/J(\l)$ and the UV one is  $2J(\l)$. It turns out that for $\l > 0$, $J(\l)$ decreases as a function of $\l$ and so the density of states becomes more and more peaked. The deformed correlator will then approach its UV form. For $\l < 0$, the situation is reversed. The density of states becomes more spread and the correlator approaches its IR form. 
The generalization to other values of $q$ is straightforward, once $c$ is known.

\section{Schwarzian theory}\label{Schwarzian}
We can also consider the application of our formulas to the Schwarzian limit of the SYK model. This is the case where we understand the bulk dual, which is just JT gravity in AdS$_2$ with Dirichlet conditions for the dilaton and the metric at the AdS boundary. As discussed in section \ref{bulk}, our deformations are changing the boundary conditions of the metric. The simplicity of the bulk theory on a disk topology leads to the calculability of higher genus corrections and a random matrix interpretation of the boundary theory \cite{Saad:2019lba}. As discussed in \cite{Stanford:2019vob}, there are two pieces of data needed to determine the random matrix model. The first is the symmetry class of matrices that one integrates over in the integral, which is set by the bulk theory. In the case of JT gravity on orientable surfaces, one has a GUE-like matrix theory. Let us stick to this case for simplicity. The other piece of data is the potential, which is determined by the spectral curve, which in turn is determined by $\rho_0(E)$ in $Z(\b) = \int dE e^{-\beta E}\rho_0(E)$, where $Z(\b)$ is computed with a disk topology. Given the simple closed-form expression \eqref{spectralcurve} for the change in $\rho_0(E)$, we can compute the new spectral curve, which determines (implicitly) the potential of the matrix integral. As the symmetry class we integrate over remains fixed, we have all the data needed for the new matrix model. In certain cases we can even compute the new partition function. For example, consider the $1d$ \ttbar deformation. The JT and super-JT theory have the following disk partition function and $\rho_0(E)$:
\begin{align}
Z_{JT}(\b) = \f{1}{4\sqrt{\pi}\b^{3/2}}e^{\pi^2/\b},\,\qquad \rho_0(E) = \f{\sinh (2\pi \sqrt{E})}{4\pi^2}\\
 \qquad Z_{SJT} = \sqrt{\f{2}{\pi \b}} \,e^{\pi^2/\b}\,,\qquad \rho_0(E) = \f{\sqrt{2}\cosh(2\pi \sqrt{E})}{\pi\sqrt{E}}
\end{align}
where we used the normalization of \cite{Stanford:2019vob}. Let us denote these partition functions collectively as $Z_n(\b) = \frac{a_n}{\b^n}e^{b_n/\b}$, where $n = 3/2, 1/2$ refer to the JT and super-JT partition functions, respectively. Upon deformation, using the kernel \eqref{generalizedKernel}, we get 
\be
Z_n(\b)_{\l} = a_n \left(\frac{c_2}{\b^2 + 8 b_n \l}\right)^{\frac{2n+1}{4}}\frac{\b \exp\left(-\frac{c_1\b}{4\l}\right) K_{n+1/2}\left( -\frac{\sqrt{c_2\b^2 + c_2 8 b_n \l}}{4\l} \right)}{\sqrt{-2\pi \l}}\,.
\ee
The density of states for $n=3/2$, which corresponds to the case of JT gravity, becomes  
\be
\rho_{\l}(E) = \frac{1}{4\pi^2}(c_1 - 4 E \l) \sinh\left(2\pi \sqrt{ \frac{c_2 - (c_1 - 4 \l E)^2}{8\l} }\right).
\ee
While the computation of these partition functions is not strictly necessary for determining the potential of the new matrix integral, which is just a function of  $\rho_\l(E)$, it suggests that the bulk path integrals can be carried out and checked against the random matrix predictions. 

We can also transform what is known as the ``trumpet" partition function \cite{Saad:2019lba}, or the partition function of the $\gamma$-Schwarzian \cite{Anninos:2018svg}. This is the path integral of a Schwarzian action over diff$(S^1)/U(1)$. The partition function is similar to the JT supergravity partition function in that the prefactor is $\b^{-1/2}$ instead of $\b^{-3/2}$ \cite{Stanford:2017thb}. We get the partition function $Z_{1/2}$, but now with $a_n = 1/\sqrt{4\pi}$ and $b_n = -\g^2/4$, instead of the super-JT values.

Another theory with a matrix integral interpretation is the $(2,p)$ minimal string theory, which has 
\be
\rho_0(E) = \f{1}{4\pi^2}\sinh\left(\f{p}{2}\,\text{arccosh}\left(1+\f{8\pi^2E}{p^2}\right)\right).
\ee
This gives the density of eigenvalues of the JT theory as $p\rightarrow \infty$ at fixed $E$. Using \eqref{spectralcurve} one can engineer a deformation $f(H)$ that turns the spectral curve for JT gravity into that of the $(2,p)$ minimal string theory. The theories, however, are different: while the $f(H)$ deformation is purely a change of boundary conditions of JT gravity, the minimal string theory has different bulk degrees of freedom. 

\subsection{General deformations and chaos}

Another interesting aspect of the Schwarzian theory to consider is its maximally chaotic behavior. We expect that the Lyapunov exponent does not change since the deformed theory still has the $SL(2,\mathbb{R})$ symmetries and no enhanced symmetry. This means that the pole structure of the momentum space correlator of fluctuations around the saddlepoint, in particular the poles at frequencies $\pm 1$, remains the same. The pole structure is directly related to the Lyapunov exponent through an analytic continuation, so the Lyapunov exponent should not change. 

Let us now verify this by explicit computation. Along the way we will see that we are able to write down an action for the deformed Schwarzian theory for an \emph{arbitrary} $f(H)$ deformation to the Hamiltonian. 

Consider the Hamiltonian of the Schwarzian theory,
\be\label{Hsch}
H = \frac{p_2^2 q_2^2}{2C} + \frac{C}{2} q_2^2 + p_1 q_2,
\ee
where we introduced two momenta $p_1, p_2$ as was done in \cite{Gross:2019ach} and $C = \Phi_r/(8\pi G)$. Let us now consider a general deformation $H \to f(H)$ with invertible $f$. We would now like to Legendre transform back to a Lagrangian, find the saddlepoints and compute the two-point function of the fluctuations around the saddlepoint. For the Legendre transformation we will also need $\dot{f}$ to be invertible. Due to the linear appearance of $p_1$ in the undeformed Hamiltonian, we can invert Hamilton's equations to solve for $p_i$ in terms of $q_i$, $q'_i$ for an arbitrary deformation $f(H)$ up to the restrictions mentioned above. The deformed Euclidean Lagrangian takes the form
\be\label{deformedSch}
L_E(\l) = \frac{C}{2} \frac{e^{\phi}}{\t'}(\phi'^2 - \t'^2) + f(\dot{f}^{-1}(e^{-\phi}\t')) - e^{-\phi}\t' \dot{f}^{-1}(e^{-\phi}\t'),
\ee
where we introduced $q_1 = \t$ and $q_2 = e^{\phi}$ and performed the required analytic continuation to Euclidean signature as in \cite{Gross:2019ach}. Here $\dot{f}(x) = \partial_x f(x)$ and $' = \partial_u$. This action is the sum of the original action\footnote{As $\l \rightarrow 0$, $f(H) \approx H + \l k H^2 + \dots$, for which the second and third term in \eqref{deformedSch} become $-(e^{-\phi}\t' - 1)^2/(4k\l)$. This enforces the constraint $e^{\phi} = \t'$ at $\l = 0$, which makes the first term equal to the undeformed Schwarzian.} and a piece involving the deformation.

It is not hard to find saddlepoints of this action. The only non-trivial saddlepoint is the one for $\phi$, since $\t(u) = u$ should remain a saddle in the deformed theory. We solve the $\phi$ equation of motion by considering an ansatz $e^{\phi} = b$, which leaves us with the constraint
\be
b^2 C - 2 \dot{f}^{-1}(1/b) = 0\,.
\ee
Since $f(H) = H + \l k H^2 + \dots$ for small $\l$ and some constant $k$, one is always guaranteed a solution that connects smoothly to the undeformed solution, $b = 1-C k \l+\dots$. Let us expand the action \eqref{deformedSch} around this solution by writing $\t(u) = u + \e(u)$, $e^{\phi} = b e^{\eta(u)}$, with $\e, \eta$ $2\pi$-periodic. Ignoring the constant piece, we find
\be
S_E(\l) = \frac{C}{2 G_f(b)} \int_0^{2\pi} du \left( -\e'(u)^2 + 2 \e'(u)\eta(u) - \eta(u)^2 + b G_f(b) (\eta'(u)^2  - \eta(u)^2) \right)
\ee
with $G_f(b) = C b^2 \ddot{f}(\dot{f}^{-1}(1/b))$. We can now easily extract the $\e$ propagator in momentum space in the usual way:
\be
\braket{\e(u)\e(0)} = \frac{1}{2\pi C b} \sum_{n \neq 0,\pm 1} \frac{1 -  b G_f(b) (n^2-1)}{n^2(n^2-1)} e^{in u}.
\ee
The poles at $n = 0,\pm 1$ from the undeformed case are still present. These poles come from the unbroken $SL(2,\mathbb{R})$ gauge symmetry and so we should not sum over $n = 0,\pm 1$. The sum above can be evaluated by writing it as a contour integral and deforming the contour so that it only encircles the poles at $n = 0,\pm 1$. This leads to 
\begin{align}
\langle \e(u) \e(0)\rangle = \f{1}{2\pi C b} (1-(1+b G_f(b))&\left(\f{\pi^2}{3}-\pi u + \f{u^2}{2}\right)+\left(\f{5}{2}+2 b G_f(b)\right)\cos u + \left.(u-\pi)\sin u \right)
\end{align}
for $0< u < 2\pi$, which is then periodically repeated. At late Lorentzian times, this has a piece that grows as $e^{2\pi t/\beta}$. Thus, it is expected that for generic $SL(2,\mathbb{R})$ invariant couplings to matter this mode will lead to a four-point matter OTOC which is maximally chaotic. For example, consider the coupling often used in the case of the undeformed Schwarzian, which gives an on-shell action $S_{\text{on-shell}} \sim \int du du' \left[\f{\t'(u)\t'(u')}{(\t(u)-\t(u'))^2}\right]^{\D} \chi(u)\chi(u')$ for a source $\chi$ of an operator of dimension $\D$. Then, by the usual procedure (see e.g. \cite{Maldacena:2016upp}), the matter four-point function is given in terms of the $\e$ two-point function. Analytically continuing and going to late time shows that the Lyapunov exponent is unchanged while the scrambling time changes according to $G_f(b)$. 

The Lyapunov exponent was found to be unchanged under the 1d \ttbar deformation in \cite{Gross:2019ach} and the 2d \ttbar deformation in \cite{He:2019vzf}. 

Let us make one final remark. As discussed already, the action \eqref{deformedSch} breaks into two pieces, with the first term being the Schwarzian action. It is tempting to  interpret the rest of the expression as a coupling  to a form of 1d gravity. For the 1d \ttbar deformation, upon replacing $e^{\phi}$ with $e^{-1}\t'$, we recover the worldline gravity theory of \cite{Gross:2019ach} in static gauge. The form \eqref{deformedSch} may provide a clue to more general worldline gravity definitions for more general deformations. 

\section{Discussion}
In this section we would like to discuss some further potential applications and speculations. 

\subsection*{\it Worldline gravity picture for correlators}
In \cite{Gross:2019ach}, we proposed a formulation of the $1d$ \ttbar deformation \eqref{holoDef} in terms of worldline gravity.  By coupling the undeformed theory to a theory of one-dimensional gravity and performing the path integral over all fields, we showed that the thermal partition function is precisely reproduced. This is analogous to the description of the \ttbar deformation of two-dimensional theories as coupling to JT gravity or 2d non-critical string theory \cite{Dubovsky:2017cnj, Dubovsky:2018bmo, Callebaut:2019omt}. 

How about correlation functions? In this paper we have given integral transform expressions for general correlation functions, so it is natural to try to generalize the worldline gravity picture to capture correlation functions as well. In fact, the integrals over the kernels that transform the correlation functions look similar to vertex operator integrals over the worldsheet in string theory, except in this case they would come with a measure for the integration provided by the kernel. So for a thermal two-point function we would want something like 
\be
\underbrace{\int \frac{\mathcal{D}e\mathcal{D}X \mathcal{D}\Phi}{\rm Vol(\rm Diff)} e^{-S[e,X;\l] -S_0[e,\Phi] }}_{\text{gravitational path integral}}\underbrace{\int e d\t' K(\t,\t')O(\t')O(0)}_{\text{vertex op integral over worldgraph}}.
\ee
This would also delocalize the operators an amount set by $\lambda$, fitting with a gravitational interpretation of the theory. But something like this does not immediately follow from our integral transforms since the kernels are integrated against the correlator itself, not the operators. This means that part of the path integral -- the part over the fields $\Phi$ of the original theory -- needs to be performed before the kernel can be integrated over. A more involved procedure, potentially by dressing the operators of the undeformed theory with the operator $X$ of the gravitational theory, seems required.

Another way to proceed is as follows. Recall that the 1d gravity action is given as 
\be\label{worldlineQG}
S = -\f{1}{8\l}\int_0^{\b'} d\t e \left(e^{-1} \dot{X}-1\right)^2\,.
\ee
Instead of taking periodic boundary conditions on $X$ as is necessary for the thermal partition function, one could consider Dirichlet boundary conditions $X(0) = X_1$ and $X(\b') = X_2$. The closed worldline thus becomes an open one with ``boundary" states $\ket{X_1}$ and $\ket{X_2}$ at its ends. This is similar to the D-brane boundary condition proposed for $T\bar{T}$ deformed 2d CFTs in \cite{Callebaut:2019omt}. The path integral we then wish to compute is a transition amplitude:
\be\label{Gprop}
G(X_1,X_2) = \int_{X(0) = X_1}^{X(\b') = X_2} \frac{\mathcal{D}e \mathcal{D}X \mathcal{D}\Phi}{\vol(\rm Diff)} e^{-S[e,X;\l] - S_0[e,\Phi]}.
\ee
The explicit computation is analogous to the one presented in \cite{Gross:2019ach} and can be shown to not quite yield the correct integral transform we presented in section \ref{correlators}. Furthermore, taking the $\l\rightarrow 0$  limit does not yield an observable in the undeformed theory, even though in the deformed theory (i.e. the worldline) it is a perfectly well-defined observable. Finally, the fields $\Phi$ of the initial theory have not made their appearance in $G$. 

To interpret our results in section \ref{correlators} in terms of a transition amplitude, we therefore need to alter the path integral \eqref{Gprop}. The operator insertions are straightforward to deal with; the tricky part is to change the integral transform. To see how it should be changed, it is convenient to Fourier transform the integral transform \eqref{holodeform} for the correlator $G$ to momentum space. It can then be seen that the integral transform can be obtained from the transition amplitude \eqref{Gprop} by certain insertions of momentum $p$. This  looks like an unnatural observable from the worldline perspective and was engineered to give the answer we wanted.  We leave a detailed study of correlation functions from the worldline gravity perspective to future work.

\subsection*{\it Coupling to other $1d$ gravities}

In the above and \cite{Gross:2019ach}, we considered couplings of the original quantum mechanics to a worldline, like \eqref{worldlinetheory}. One might wonder whether there exist other couplings to one-dimensional gravity that could also be interpreted as a deformation of the form $H \to f(H)$. One obvious candidate is a covariant version of the Schwarzian theory, or its cousin the $\g$-Schwarzian. For the usual Schwarzian theory there does not seem to be a clean interpretation in terms of $H \to f(H)$, since it introduces additional prefactors to the Boltzmann weights. One could interpret these addditional prefactors as changes in the energy eigenstates, hence bringing us outside the deformations studied here. 
For the $\g$-Schwarzian with the periodicity of the Schwarzian field fixed to one (or equivalently setting $b=1$ in the trumpet geometry), there are no such additional prefactors and its coupling to the initial quantum mechanics can be interpreted as a deformation sending the original energies $E$ to $\sqrt{E}$.

\subsection*{\it Coupling multiple systems}
An interesting application of the more general $f(H, Q_i)$ deformations discussed in section \ref{intro} occurs in the case where the additional charges $Q_i$ are Hamiltonians of independent systems. For example, take $n$ decoupled systems with Hamiltonians $H_1, \dots, H_n$. The Hamiltonian of the full system is $H_1 + H_2 + \dots + H_n$. Consider a deformation to this Hamiltonian of the form $H\rightarrow f(H_i)$. This can introduce couplings between the independent systems, although interestingly we still have $n$ conserved charges, corresponding to the original Hamiltonians. A simple example is $n=2$ with $f(H_1, H_2) = H_1 + H_2 + \lambda H_1 H_2$. In a theory of fermions like SYK the interaction term will not introduce higher derivatives. It would be interesting to explore applications of such couplings, where by the arguments in this paper the observables are easily calculable. A natural setup is the context of spin chains, where recent work has focused on the two-dimensional \ttbar deformation to the spin chain \cite{Marchetto:2019yyt, Pozsgay:2019ekd}.

\section*{Acknowledgments}
It is a pleasure to thank Tarek Anous, Louise Anderson, Matt Headrick, Luca Iliesiu, Igor Klebanov, Edward Mazenc, G\'abor S\'arosi, Eva Silverstein, Ronak Soni, Douglas Stanford, Zhenbin Yang and Peng Fei Zhang for enlightening discussions. The research of JK is supported by the Simons Foundation. The work of ES is supported by the Simons Foundation as part of the Simons Collaboration on the Nonperturbative Bootstrap. AR is supported by the National Science Foundation under Grant No. NSF PHY-1748958, the Department of Energy under DE-SC0009987, and by the Simons Foundation through the It from Qubit Simons Collaboration on Quantum Fields, Gravity and Information. DG is supported by NSF grant 1125915. ES would like to thank the Aspen Center for Physics, which is supported by National Science Foundation under grant PHY-1607611, and the Kavli Institute for Theoretical Physics, which is supported by the National Science Foundation under grant PHY-1748958, where parts of this work were completed. ES and JK would also like to thank the Aruba workshop ``Qubits on the Horizon" where part of this work was completed.

\appendix

\section{Deformations with explicit kernels}\label{app:Kernels}

In this appendix we give a few examples for which one can find explicit kernels $K_f(\b,\b')$. For instance, we can generalize the $1d$ \ttbar deformation to  
\be
f(H) = \frac{1}{4\l}\left( c_1(\l) - \sqrt{c_2(\l)-8\l H} \right),
\ee
with $c_i(\l)$ arbitrary functions of $\l$ for which the kernel becomes
\be\label{generalizedKernel}
K_f(\b,\b') = \frac{\b}{\b'^{3/2}\sqrt{-8\pi \l}} \exp\left( \frac{(\b^2 - 2 c_1(\l) \b \b' + c_2(\l) \b'^2)}{8\b' \l}\right).
\ee
Here $\l < 0$. The contour $\mathcal{C}$ is again running from $0$ to $\infty$. These types of deformations arise for example when considering general dilaton gravities at finite cutoff or the one-dimensional analogue of the $\Lambda_2$-flow, the $\Lambda_1$-flow for which $c_2 = -1$ and $c_1$ set by the choice of counterterm action (although the dS$_2$ flow is for $\l > 0$) \cite{Gorbenko:2018oov, Gross:2019ach}. For general $c_i$ this kernel does not have a well-defined limit as $\l \to 0$. Furthermore, covergence of the integral transform depends on the sign of $c_2$ (in particular for ordinary $Z(\b')$ we need $c_2 > 0$ to have exponential suppression at large $\b'$). In the cases where the kernel is well-defined, it can be given a worldline gravity interpretation. In particular, the deformation above leads to a thermal partition function that can be reproduced by coupling the original theory to a theory of 1d gravity given by the unit winding sector of
\be\label{worldlinetheory}
S = -\f{1}{8\l}\int_0^{\b'} d\t e \left(e^{-1} \dot{X}^2-2 c_1 \dot{X}+c_2 e\right)\,.
\ee
The explicit steps are exactly analogous to \cite{Gross:2019ach}.

The inverse of the above deformation can also be found straightforwardly and gives the kernel
\be
K_f(\b,\b') = \frac{1}{\sqrt{8\pi \l \b}}\exp\left(-\frac{\b'^2 - 2c_1(\l) \b \b' + c_2(\l)\b^2}{8\l \b} \right).
\ee

There are other cases where explicit kernels exist, but it is not our intention to provide an exhaustive list. A simple family consists of the deformations $f(H) = H + \l H^n$ with $n$ a positive integer. These kernels are defined using a Fourier transform of $e^{-\b f(E)}$ and give sums of hypergeometric functions. The partition function is obtained via the inverse Fourier transform, which requires integrating the undeformed partition function over imaginary temperatures. 

Further explicit kernels can be generated by iteration; given functions $f_i(E)$ with known kernels $K_{f_i}$, one can carry out integral transforms multiple times to treat deformations of the form $f_1(f_2(...f_n(E)))$.

\section{Numerical implementation of integral transforms}\label{numerics}
\begin{figure}
    \centering
    \includegraphics{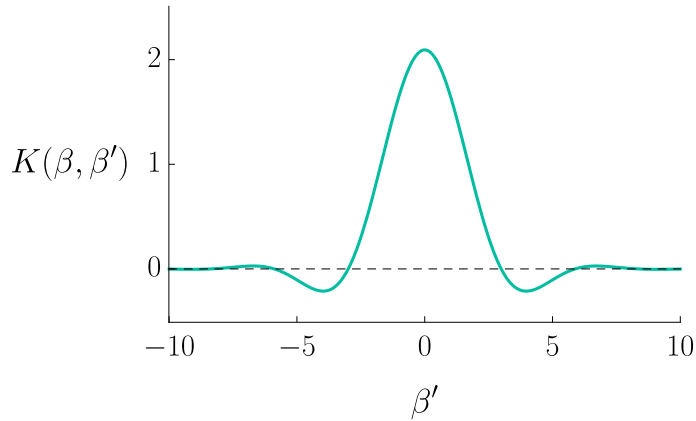}
    \caption{Numerically computed kernel $K(\b,\b')$ for the deformation $f(H) = H + \l (H^2 + H^4)$ for $\b = 1$ and $\l = 0.5$. } 
    \label{fig:Kernel}
\end{figure}

\begin{figure}
    \centering
    \includegraphics{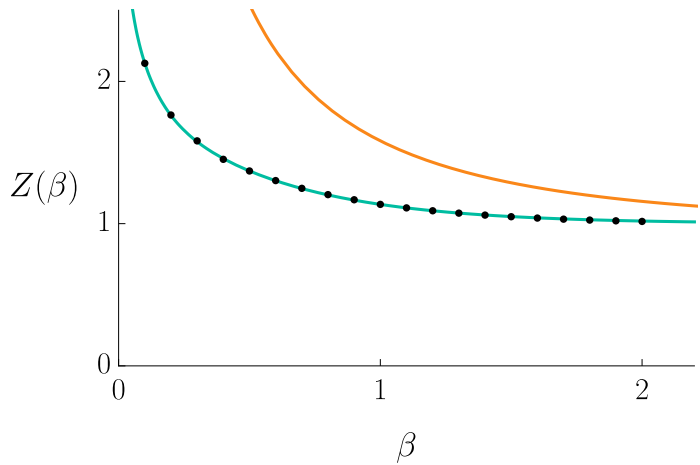}
    \caption{Solid orange line: The undeformed partition function  for the simple harmonic oscillator ($\w = 1$). Solid aqua blue line: Truncated sum (to 40 terms) of the partition function for the deformed simple harmonic oscillator for the deformation $f(H) = H + \l (H^2 + H^4)$  with $\l = 0.5$. Dots: Numerically computed deformed partition function through the integral transform.} 
    \label{fig:Z}
\end{figure}

In this section we will give a simple numerical application of the integral transforms discussed in section \ref{correlators}. Say we have a kernel defined as
\be
e^{-\b f(E)} = \int_{-\infty}^{\infty} d\b' e^{-i \b' E}K_f(\b,\b'),
\ee
so that we can get $K_f$ using a Fourier tranform
\be
K_f(\b,\b') = \int_{-\infty}^{\infty} \frac{dE}{2\pi} e^{i \b' E} e^{-\b f(E)}\,.
\ee
Let's consider the deformation $f(H) = H + \l (H^2 + H^4)$, for which we do not have a closed-form expression for the kernel. Numerically computing the kernel gives us figure  \ref{fig:Kernel}. 
The partition function of the deformed theory is then
\be
Z_{\l}(\b) = \int_{-\infty-i \e}^{\infty-i \e} d\b' K_f(\b,\b') Z(i\b')\,,
\ee
with $\e>0$ added for convergence.
The exact deformed partition function is given by
\be
Z_\lambda(\beta) = \sum_E e^{-\beta f(E)}\,.
\ee
The two are compared in figure \ref{fig:Z}. 

\section{Symmetries}

An important feature of the $f(H)$ deformations is that all conserved charges in the original theory remain conserved in the deformed theory \cite{Gross:2019ach}. Another interesting possibility to consider is that of enhanced symmetry in the deformed theory. For invertible $f(H)$, this is not possible because we can apply our argument that all charges are preserved in the deformed theory to the deformation $f^{-1}(H)$.

We can also discuss the possibility of spontaneous symmetry breaking (e.g. in large-$N$ systems). Picking a strictly monotonic $f(H)$ means the symmetry spontaneously breaks in the deformed theory if and only if it spontaneously breaks in the undeformed theory. In particular, the symmetry is not broken if $Q|\psi\rangle = 0$ for charge operator $Q$ and ground state $|\psi\rangle$. Since we diagonalize our deformed system by the same set of eigenstates and pick a strictly monotonic $f(H)$, the vacuum of the deformed theory is exactly the vacuum of the undeformed theory, and it will again be annihilated by the charge operator $Q$. The converse can be proven in the same way: take $Q|\psi\rangle = 0$ in the deformed theory, consider the deformation $f^{-1}(H)$, and apply the argument from before. In some instances the conserved charge of the deformed theory may be written slightly differently. For example, the deformed Hamiltonian or deformed supercharges are not simply the $H$ and $Q_i$ of the original theory, but in any such case they can still be written as some function of the charges of the original theory, so the argument above again applies. 

The argument above tells us that we should expect replica symmetry to be preserved in the deformed SYK models considered in section \ref{SYK} since it appears to be preserved in the original SYK model \cite{Maldacena:2016hyu, Bagrets:2016cdf, Gur-Ari:2018okm, Garcia-Garcia:2016mno, Cotler:2016fpe, Arefeva:2018vfp, PhysRevB.63.134406 , PhysRevB.94.035135}. 


\small
\bibliographystyle{ourbst}
\bibliography{1dBib}

\providecommand{\href}[2]{#2}\begingroup\raggedright\begin{thebibliography}{10}

\bibitem{Anninos:2017hhn}
D.~Anninos and D.~M. Hofman, ``{Infrared Realization of dS$_2$ in AdS$_2$},''
  \href{http://dx.doi.org/10.1088/1361-6382/aab143}{{\em Class. Quant. Grav.}
  {\bfseries 35} no.~8, (2018) 085003},
\href{http://arxiv.org/abs/1703.04622}{{\ttfamily arXiv:1703.04622 [hep-th]}}.

\bibitem{Gross:2019ach}
D.~J. Gross, J.~Kruthoff, A.~Rolph, and E.~Shaghoulian, ``{$T\overline{T}$ in
  AdS$_2$ and Quantum Mechanics},''
\href{http://arxiv.org/abs/1907.04873}{{\ttfamily arXiv:1907.04873 [hep-th]}}.

\bibitem{Dirac:1938nz}
P.~A.~M. Dirac, ``{Classical theory of radiating electrons},''
\href{http://dx.doi.org/10.1098/rspa.1938.0124}{{\em Proc. Roy. Soc. Lond.}
  {\bfseries A167} (1938) 148--169}.

\bibitem{Bhabha:1946zz}
H.~J. Bhabha, ``{On the Expansibility of Solutions in Powers of the Interaction
  Constants},''
\href{http://dx.doi.org/10.1103/PhysRev.70.759}{{\em Phys. Rev.} {\bfseries 70}
  (1946) 759--760}.

\bibitem{Cavaglia:2016oda}
A.~Cavagli\`{a}, S.~Negro, I.~M. Sz\'{e}cs\'{e}nyi, and R.~Tateo, ``{$T
  \bar{T}$-deformed 2D Quantum Field Theories},''
  \href{http://dx.doi.org/10.1007/JHEP10(2016)112}{{\em JHEP} {\bfseries 10}
  (2016) 112},
\href{http://arxiv.org/abs/1608.05534}{{\ttfamily arXiv:1608.05534 [hep-th]}}.

\bibitem{Smirnov:2016lqw}
F.~A. Smirnov and A.~B. Zamolodchikov, ``{On space of integrable quantum field
  theories},'' \href{http://dx.doi.org/10.1016/j.nuclphysb.2016.12.014}{{\em
  Nucl. Phys.} {\bfseries B915} (2017) 363--383},
\href{http://arxiv.org/abs/1608.05499}{{\ttfamily arXiv:1608.05499 [hep-th]}}.

\bibitem{Guica:2019nzm}
M.~Guica and R.~Monten, ``{$T\bar T$ and the mirage of a bulk cutoff},''
\href{http://arxiv.org/abs/1906.11251}{{\ttfamily arXiv:1906.11251 [hep-th]}}.

\bibitem{Cooper:1994eh}
F.~Cooper, A.~Khare, and U.~Sukhatme, ``{Supersymmetry and quantum
  mechanics},'' \href{http://dx.doi.org/10.1016/0370-1573(94)00080-M}{{\em
  Phys. Rept.} {\bfseries 251} (1995) 267--385},
\href{http://arxiv.org/abs/hep-th/9405029}{{\ttfamily arXiv:hep-th/9405029
  [hep-th]}}.

\bibitem{Grassi:2018bci}
A.~Grassi and M.~Mari\~{n}o, ``{A Solvable Deformation of Quantum Mechanics},''
  \href{http://dx.doi.org/10.3842/SIGMA.2019.025}{{\em SIGMA} {\bfseries 15}
  (2019) 025},
\href{http://arxiv.org/abs/1806.01407}{{\ttfamily arXiv:1806.01407 [hep-th]}}.

\bibitem{deAlfaro:1976vlx}
V.~de~Alfaro, S.~Fubini, and G.~Furlan, ``{Conformal Invariance in Quantum
  Mechanics},''
\href{http://dx.doi.org/10.1007/BF02785666}{{\em Nuovo Cim.} {\bfseries A34}
  (1976) 569}.

\bibitem{Chamon:2011xk}
C.~Chamon, R.~Jackiw, S.-Y. Pi, and L.~Santos, ``{Conformal quantum mechanics
  as the CFT$_1$ dual to AdS$_2$},''
  \href{http://dx.doi.org/10.1016/j.physletb.2011.06.023}{{\em Phys. Lett.}
  {\bfseries B701} (2011) 503--507},
\href{http://arxiv.org/abs/1106.0726}{{\ttfamily arXiv:1106.0726 [hep-th]}}.

\bibitem{Cardy:2019qao}
J.~Cardy, ``{$T\overline T$ deformation of correlation functions},''
\href{http://arxiv.org/abs/1907.03394}{{\ttfamily arXiv:1907.03394 [hep-th]}}.

\bibitem{Hartman:2018tkw}
T.~Hartman, J.~Kruthoff, E.~Shaghoulian, and A.~Tajdini, ``{Holography at
  finite cutoff with a $T^2$ deformation},''
  \href{http://dx.doi.org/10.1007/JHEP03(2019)004}{{\em JHEP} {\bfseries 03}
  (2019) 004},
\href{http://arxiv.org/abs/1807.11401}{{\ttfamily arXiv:1807.11401 [hep-th]}}.

\bibitem{kitaevKITP}
A.~Kitaev, ``\href{http://online.kitp.ucsb.edu/online/entangled15/}{A simple
  model of quantum holography, Part 1 $\&$ 2, talk given at the Entanglement in
  Strongly-Correlated Quantum Matter, KITP, University of California, Santa
  Barbara, California, U.S.A.},''.

\bibitem{PhysRevLett.70.3339}
S.~Sachdev and J.~Ye, ``Gapless spin-fluid ground state in a random quantum
  heisenberg magnet,''
  \href{https://link.aps.org/doi/10.1103/PhysRevLett.70.3339}{{\em Phys. Rev.
  Lett.} {\bfseries 70} (May, 1993) 3339--3342}.

\bibitem{Herzog:2010hf}
C.~P. Herzog, I.~R. Klebanov, S.~S. Pufu, and T.~Tesileanu, ``{Multi-Matrix
  Models and Tri-Sasaki Einstein Spaces},''
  \href{http://dx.doi.org/10.1103/PhysRevD.83.046001}{{\em Phys. Rev.}
  {\bfseries D83} (2011) 046001},
\href{http://arxiv.org/abs/1011.5487}{{\ttfamily arXiv:1011.5487 [hep-th]}}.

\bibitem{10.2307/1970079}
E.~P. Wigner, ``Characteristic vectors of bordered matrices with infinite
  dimensions,'' \href{http://www.jstor.org/stable/1970079}{{\em Annals of
  Mathematics} {\bfseries 62} no.~3, (1955) 548--564}.

\bibitem{Saad:2019lba}
P.~Saad, S.~H. Shenker, and D.~Stanford, ``{JT gravity as a matrix integral},''
\href{http://arxiv.org/abs/1903.11115}{{\ttfamily arXiv:1903.11115 [hep-th]}}.

\bibitem{Stanford:2019vob}
D.~Stanford and E.~Witten, ``{JT Gravity and the Ensembles of Random Matrix
  Theory},''
\href{http://arxiv.org/abs/1907.03363}{{\ttfamily arXiv:1907.03363 [hep-th]}}.

\bibitem{Anninos:2018svg}
D.~Anninos, D.~A. Galante, and D.~M. Hofman, ``{De Sitter Horizons \&
  Holographic Liquids},'' \href{http://dx.doi.org/10.1007/JHEP07(2019)038}{{\em
  JHEP} {\bfseries 07} (2019) 038},
\href{http://arxiv.org/abs/1811.08153}{{\ttfamily arXiv:1811.08153 [hep-th]}}.

\bibitem{Stanford:2017thb}
D.~Stanford and E.~Witten, ``{Fermionic Localization of the Schwarzian
  Theory},'' \href{http://dx.doi.org/10.1007/JHEP10(2017)008}{{\em JHEP}
  {\bfseries 10} (2017) 008},
\href{http://arxiv.org/abs/1703.04612}{{\ttfamily arXiv:1703.04612 [hep-th]}}.

\bibitem{Maldacena:2016upp}
J.~Maldacena, D.~Stanford, and Z.~Yang, ``{Conformal symmetry and its breaking
  in two dimensional Nearly Anti-de-Sitter space},''
  \href{http://dx.doi.org/10.1093/ptep/ptw124}{{\em PTEP} {\bfseries 2016}
  no.~12, (2016) 12C104},
\href{http://arxiv.org/abs/1606.01857}{{\ttfamily arXiv:1606.01857 [hep-th]}}.

\bibitem{He:2019vzf}
S.~He and H.~Shu, ``{Correlation functions, entanglement and chaos in the
  $T\bar{T}$/$J\bar{T}$-deformed CFTs},''
\href{http://arxiv.org/abs/1907.12603}{{\ttfamily arXiv:1907.12603 [hep-th]}}.

\bibitem{Dubovsky:2017cnj}
S.~Dubovsky, V.~Gorbenko, and M.~Mirbabayi, ``{Asymptotic fragility, near
  AdS$_{2}$ holography and $ T\overline{T} $},''
  \href{http://dx.doi.org/10.1007/JHEP09(2017)136}{{\em JHEP} {\bfseries 09}
  (2017) 136},
\href{http://arxiv.org/abs/1706.06604}{{\ttfamily arXiv:1706.06604 [hep-th]}}.

\bibitem{Dubovsky:2018bmo}
S.~Dubovsky, V.~Gorbenko, and G.~Hern\'{a}ndez-Chifflet, ``{$ T\overline{T} $
  partition function from topological gravity},''
  \href{http://dx.doi.org/10.1007/JHEP09(2018)158}{{\em JHEP} {\bfseries 09}
  (2018) 158},
\href{http://arxiv.org/abs/1805.07386}{{\ttfamily arXiv:1805.07386 [hep-th]}}.

\bibitem{Callebaut:2019omt}
N.~Callebaut, J.~Kruthoff, and H.~Verlinde, ``{$T\bar{T}$ deformed CFT as a
  non-critical string},''
\href{http://arxiv.org/abs/1910.13578}{{\ttfamily arXiv:1910.13578 [hep-th]}}.

\bibitem{Marchetto:2019yyt}
E.~Marchetto, A.~Sfondrini, and Z.~Yang, ``{$T\bar{T}$ deformations and
  integrable spin chains},''
\href{http://arxiv.org/abs/1911.12315}{{\ttfamily arXiv:1911.12315 [hep-th]}}.

\bibitem{Pozsgay:2019ekd}
B.~Pozsgay, Y.~Jiang, and G.~Takács, ``{$T\bar T$-deformation and long range
  spin chains},''
\href{http://arxiv.org/abs/1911.11118}{{\ttfamily arXiv:1911.11118 [hep-th]}}.

\bibitem{Gorbenko:2018oov}
V.~Gorbenko, E.~Silverstein, and G.~Torroba, ``{dS/dS and $ T\overline{T} $},''
  \href{http://dx.doi.org/10.1007/JHEP03(2019)085}{{\em JHEP} {\bfseries 03}
  (2019) 085},
\href{http://arxiv.org/abs/1811.07965}{{\ttfamily arXiv:1811.07965 [hep-th]}}.

\bibitem{Maldacena:2016hyu}
J.~Maldacena and D.~Stanford, ``{Remarks on the Sachdev-Ye-Kitaev model},''
  \href{http://dx.doi.org/10.1103/PhysRevD.94.106002}{{\em Phys. Rev.}
  {\bfseries D94} no.~10, (2016) 106002},
\href{http://arxiv.org/abs/1604.07818}{{\ttfamily arXiv:1604.07818 [hep-th]}}.

\bibitem{Bagrets:2016cdf}
D.~Bagrets, A.~Altland, and A.~Kamenev, ``{Sachdev–Ye–Kitaev model as
  Liouville quantum mechanics},''
  \href{http://dx.doi.org/10.1016/j.nuclphysb.2016.08.002}{{\em Nucl. Phys.}
  {\bfseries B911} (2016) 191--205},
\href{http://arxiv.org/abs/1607.00694}{{\ttfamily arXiv:1607.00694
  [cond-mat.str-el]}}.

\bibitem{Gur-Ari:2018okm}
G.~Gur-Ari, R.~Mahajan, and A.~Vaezi, ``{Does the SYK model have a spin glass
  phase?},'' \href{http://dx.doi.org/10.1007/JHEP11(2018)070}{{\em JHEP}
  {\bfseries 11} (2018) 070},
\href{http://arxiv.org/abs/1806.10145}{{\ttfamily arXiv:1806.10145 [hep-th]}}.

\bibitem{Garcia-Garcia:2016mno}
A.~M. García-García and J.~J.~M. Verbaarschot, ``{Spectral and thermodynamic
  properties of the Sachdev-Ye-Kitaev model},''
  \href{http://dx.doi.org/10.1103/PhysRevD.94.126010}{{\em Phys. Rev.}
  {\bfseries D94} no.~12, (2016) 126010},
\href{http://arxiv.org/abs/1610.03816}{{\ttfamily arXiv:1610.03816 [hep-th]}}.

\bibitem{Cotler:2016fpe}
J.~S. Cotler, G.~Gur-Ari, M.~Hanada, J.~Polchinski, P.~Saad, S.~H. Shenker,
  D.~Stanford, A.~Streicher, and M.~Tezuka, ``{Black Holes and Random
  Matrices},'' \href{http://dx.doi.org/10.1007/JHEP09(2018)002,
  10.1007/JHEP05(2017)118}{{\em JHEP} {\bfseries 05} (2017) 118},
  \href{http://arxiv.org/abs/1611.04650}{{\ttfamily arXiv:1611.04650
  [hep-th]}}.
[Erratum: JHEP09,002(2018)].

\bibitem{Arefeva:2018vfp}
I.~Aref'eva, M.~Khramtsov, M.~Tikhanovskaya, and I.~Volovich,
  ``{Replica-nondiagonal solutions in the SYK model},''
  \href{http://dx.doi.org/10.1007/JHEP07(2019)113}{{\em JHEP} {\bfseries 07}
  (2019) 113},
\href{http://arxiv.org/abs/1811.04831}{{\ttfamily arXiv:1811.04831 [hep-th]}}.

\bibitem{PhysRevB.63.134406}
A.~Georges, O.~Parcollet, and S.~Sachdev, ``Quantum fluctuations of a nearly
  critical heisenberg spin glass,''
  \href{https://link.aps.org/doi/10.1103/PhysRevB.63.134406}{{\em Phys. Rev. B}
  {\bfseries 63} (Mar, 2001) 134406}.

\bibitem{PhysRevB.94.035135}
W.~Fu and S.~Sachdev, ``Numerical study of fermion and boson models with
  infinite-range random interactions,''
  \href{https://link.aps.org/doi/10.1103/PhysRevB.94.035135}{{\em Phys. Rev. B}
  {\bfseries 94} (Jul, 2016) 035135}.

\end{thebibliography}\endgroup

\end{document}